\documentclass[useAMS,usenatbib]{mn2e}
\usepackage{amsmath}
\usepackage{txfonts}
\usepackage{graphicx}
\usepackage{textcomp}
\usepackage{natbib}
\usepackage{deluxetable}
\usepackage{mathrsfs} 

\newcommand{\um}{\textmu m }
\newcommand{\uu}{\textmu m}

\newcommand{\tausil}{$\rm{\tau_{9.7}}$ }
\newcommand{\tausilu}{$\rm{\tau_{9.7}}$}

\newcommand{\zspec}{z_{\rm{spec}} }
\newcommand{\zspecu}{z_{\rm{spec}}}

\newcommand{\zIRS}{z_{\rm{IRS}} }
\newcommand{\zIRSu}{z_{\rm{IRS}}}
\newcommand{\zbest}{z_{\rm{best}} }
\newcommand{\zbestu}{z_{\rm{best}}}
\newcommand{\zcolors}{$z${}\rm{COLORS }}
\newcommand{\zcolorsu}{$z${}\rm{COLORS}}

\bibpunct[]{(}{)}{;}{a}{}{,}

\title[Redshift measurement in mid-infrared spectra]{Automated measurement of redshift from mid-infrared low resolution spectroscopy}

\author[Hern\'an-Caballero]{
Antonio Hern\'an-Caballero\\
Instituto de F\'isica de Cantabria, CSIC-UC, Avenida de los Castros s/n, 39005, Santander, Spain.\hspace{0.3cm}E-mail: ahernan@ifca.unican.es}

\begin{document}
\date{Accepted ........ Received ........;}

\pagerange{\pageref{firstpage}--\pageref{lastpage}} \pubyear{2012}

\maketitle

\label{firstpage}

\begin{abstract}
Obtaining accurate redshifts from mid-infrared (MIR) low-resolution (R $\sim$ 100) spectroscopy is challenging because the wavelength resolution is too low to detect narrow lines in most cases. Yet, the number of degrees of freedom and diversity of spectral features are too high for regular spectral energy distribution (SED) fitting techniques to be convenient.
Here we present a new SED-fitting based routine for redshift determination that is optimised for MIR low-resolution spectroscopy. Its flexible template scaling increases the sensitivity to slope changes and small scale features in the spectrum, while a new selection algorithm called Maximum Combined Pseudo-Likelihood (MCPL) provides increased accuracy and a lower number of outliers compared to the standard maximum-likelihood (ML) approach.
Unlike ML, MCPL searches for local (instead of absolute) maxima of a ``pseudo-likelihood'' (PL) function, and combines results obtained for all the templates in the library to weed out spurious redshift solutions.
The capabilities of MCPL are demonstrated by comparing its redshift estimates to those of regular ML and to the optical spectroscopic redshifts of a sample of 491 \textit{Spitzer}/IRS spectra from extragalactic sources at 0 $<$ $z$ $<$ 3.7.
MCPL achieves a redshift accuracy $\Delta(z)/(1+z)$ $<$ 0.005 for 78\% of the galaxies in the sample compared to 68\% for ML. The rate of outliers ($\Delta(z)/(1+z)$ $>$ 0.02) is 14\% for MCPL and 22\% for ML.
$\chi^2$ values for ML solutions are found to correlate with the SNR of the spectra, but not with redshift accuracy. By contrast, the peak value of the normalised combined PL ($\gamma$) is found to provide a good indication on the reliability of the MCPL solution for individual sources.
The accuracy and reliability of the redshifts depends strongly on the MIR SED. Sources with significant polycyclic aromatic hydrocarbon emission obtain much better results compared to sources dominated by AGN continuum. 
Nevertheless, for a given $\gamma$ the frequency of accurate solutions and outliers is largely independent on their SED type.
This reliability indicator for MCPL solutions allows to select subsamples with highly reliable redshifts. In particular, a $\gamma$ $>$ 0.15 threshold retains 79\% of the sources with $\Delta(z)/(1+z)$ $<$ 0.005 while reducing the outlier rate to 3.8\%.

\end{abstract}


\section{Introduction} 

Finding a galaxy's redshift typically requires the identification of narrow emission or absorption lines in a medium- or high-resolution spectrum (spectroscopic redshift).
Alternatively, broad features of the spectral energy distribution (SED) are revealed by multi-wavelength photometry, and can be used to obtain photometric redshifts.
Spectroscopic redshifts are accurate but very time-consuming, while photometric redshifts offer limited accuracy (typically in the $\Delta z/(1+z)$ $\sim$ 0.01--0.1 range) and suffer from color degeneracies that may lead to catastrophic errors \citep{Fernandez-Soto99, Benitez00}.

Halfway between the two is low resolution spectroscopy (LRS), which at R $\sim$ 50--100 provides a compromise between sensitivity and spectral resolution. LRS has become common in space based infrared missions, particularly \textit{Spitzer} with the Infrared Spectrograph \citep[IRS;][]{Houck04}, and later AKARI and its Near-infrared Camera/spectrometer (IRC) \citep{Onaka07,Murakami07}. 

Future infrared missions will also provide LRS capabilities, 
including the Mid-InfraRed Instrument onboard JWST \citep[MIRI;][]{Wright08} and two instruments for SPICA: the Mid-InfRAred Camera w/wo LEns \citep[MIRACLE;][]{Wada10} and the far-infrared instrument \citep[SAFARI;][]{Goicoechea11}.

Since the spectrum is spread over a smaller number of pixels, mid-infrared (MIR) 5--35 \um LRS offers higher continuum sensitivity with a spectral resolution still capable of resolving many features used for diagnostics, like the polycyclic aromatic hydrocarbon (PAH) bands and absorption bands from silicates, water ice, and carbon monoxide, among others.

LRS has also the potential to yield redshifts with accuracies intermediate between those of medium resolution spectroscopy and photometric redshifts, since the theoretical redshift resolution is proportional to the wavelength resolution: $\Delta z/(1+z)$ $\sim$ $\Delta \lambda/\lambda$.
Nevertheless, narrow spectral lines are marginally unresolved at LRS resolutions, and this outweighs for them the sensitivity advantage over higher resolution spectroscopy, since the lines get washed out by the continuum emission. Because of this dilution, narrow lines are clearly detected only in high signal to noise ratio (SNR) spectra or in sources with large equivalent width (EW). Therefore, fine structure lines observed at LRS are not suitable for spectroscopic redshift determination in the general case.

In MIR spectra, the PAH and silicate bands are routinely used to estimate the redshifts of optically faint sources \citep[e.g.][]{Houck07,Yan07,Farrah09,Weedman09}. However, since they often show complex morphologies, it requires a visual inspection to properly identify them, particularly in sources with high obscuration or where a strong active galactic nucleus (AGN) continuum reduces the contrast of the features. In addition, spectra with low SNR make it difficult to identify individual features even to the trained eye.

When spectroscopic redshifts are not workable, the backup strategy is photometric redshifts.
The multiple photometric redshift techniques developed can be classified in two main groups: those based on ``learning'' with a large training set \citep[e.g.][]{Connolly95, Brunner97, Wang98, Collister04, Wadadekar05, Carliles10} and those based on ``SED-fitting'' with a set of spectral templates \citep[e.g.][]{Baum62, Koo85, Lanzetta96, Gwyn96, Sawicki97,Fernandez-Soto99,Benitez00,Bolzonella00,LeBorgne02, Babbedge04, Feldmann06}.

SED-fitting works well with broadband photometry in the optical and near-infrared (NIR) because the SED of galaxies in this range shows little diversity. In normal galaxies the SED is dominated by starlight, and it can be successfully modelled by the combination of a few stellar populations obscured by a screen of dust \citep[e.g.][]{Bruzual93,Bruzual03,Silva98} or compared to a small set of empirical \citep[e.g.][]{Assef08} or semi-empirical \citep[e.g.][]{ Coleman80, Ilbert06} templates.

Even if the galaxy hosts a low luminosity AGN or an obscured AGN of any luminosity, it has little impact on the broadband SED of the galaxy. Only quasars produce continuum emission strong enough to dominate the optical-NIR SED of the galaxy, and they become a hassle for photometric redshift routines \citep{Hatziminaoglou00, Richards01}. This is because their power-law SED does not provide high contrast features, and the broad emission lines require good coverage with narrow or intermediate band filters to be identified \citep{Benitez09,Matute12,Abramo12}

In the MIR, SED-fitting is far more problematic because the output of galaxies arises from several independent processes, including the Rayleigh-Jeans tail of stellar emission, thermal emission from hot and warm dust heated by the (active) nucleus, fluorescence of PAH molecules, radiative transitions of ionised and neutral atoms, rotational and vibrational transitions of H$_2$ and other molecules, and non-thermal radiation from radio sources (AGN, supernovae, masers). 

Furthermore, there is high dispersion in the correlation between emission from the stellar component and the interstellar medium, and population synthesis codes do not yet reproduce accurately spectral features at wavelengths $\lambda$ $\gtrsim$ 5 \um, in particular PAH emission \citep{Brodwin06}.
In practice, adding photometric points at wavelengths $\lambda$ $\gtrsim$ 5 \um to an optical-NIR SED does not improve the accuracy of the redshift solution. Nevertheless, \citet{Rowan-Robinson08} successfully apply a two-step method to fit photometry longward and shortward of 5 \um with to separate set of templates, and \citet{Negrello09} obtain $\Delta(z)/(1+z)$ $<$ 0.1 for 90\% of sources in the range 0.5 $<$ $z$ $<$ 1.5 using a combination of \textit{ISO}, \textit{Spitzer}, and \textit{AKARI} photometry in the 3.6--24 \um range.   

Template fitting can produce very accurate redshift estimates with MIR LRS if one important issue is addressed. 
Because of the diversity of MIR SEDs and the large number of data points in the spectrum (compared to photometric SEDs), 
it cannot be expected that every source in a survey will find an accurate model of its MIR SED in the template library.
As a consequence, the standard approach of SED-fitting photometric redshifts (that is, shifting and scaling of the template, and a $\chi^2$ minimisation to find the best fit) needs to be modified. This is because it favours the templates that best reproduce the overall shape of the continuum even if the smaller scale features (the ones capable of producing an accurate redshift) are poorly fit or misplaced.

A good match at certain redshift between small scale spectral features of the spectrum and a template will be signaled by a sudden drop in the value of the $\chi^2$ statistics relative to values for similar redshifts. This may not be the absolute minimum in $\chi^2$ if the shape of the continuum is somewhat different for the spectrum and template, and there may be more than one such dips if one or more features produce partial matches by chance.

In addition, it is likely that several templates have at least some features in common with the spectrum. Each of them will produce a drop in $\chi^2$ at the actual redshift of the source, while spurious alignments can occur at different redshifts for each template. Thus, filtering the redshift values at which different templates obtain local minima of $\chi^2$, and then combining them in a way that favours strong dips as well as frequently repeated redshifts, allows to obtain a redshift solution that is much more robust than finding the absolute minimum of $\chi^2$ for any template. 

In this work a routine for redshift estimation from MIR LRS based on these principles is presented, and its capabilities demonstrated using a large sample of extragalactic sources with both optical and \textit{Spitzer}/IRS spectroscopy. 

The outline of the paper is as follows. Section 2 presents the algorithm for redshift estimation and explains the features that depart from regular $\chi^2$ minimisation routines. \S3 describes the sample selection and \S4 the template library. \S5 evaluates the accuracy and reliability of the redshift estimates obtained and their dependency with the MIR type of the source. \S6 briefly summarizes the main conclusions of the paper.

\section{The method}\label{themethod}

The redshift determination algorithm described here is implemented by the \textit{Redshift COde for LOw-Resolution Spectroscopy} (\zcolorsu), developed by the author. 
It obtains the redshift of a source as well as an estimate of its reliability by comparing its MIR spectrum (hereafter, the spectrum) with a set of spectral templates (hereafter, the templates).

Let F($\lambda$) be the flux density of the spectrum as a function of the observed wavelength $\lambda$, and S$_k$($\lambda'$) the flux density of the $k$-th template as a function of the restframe wavelength $\lambda'$.

Both the spectrum F($\lambda$) and templates S$_k$($\lambda'$) are resampled to a common grid of wavelength values $\lambda_i$ = $\lambda_0 e^{\beta i}$. The parameter $\beta$ determines the spectral resolution R = 1/$\beta$ of the resampled spectrum and templates.
The uniform spacing in $\log \lambda$ is convenient for computational efficiency reasons, since such sampling ensures that the set of redshift values $z_j$ = $e^{\beta j}-1$, also evenly spaced in $\log(1+z)$, verifies $\lambda_{i+j}$ = $\lambda_i (1+z_j)$ for every \{i, j\}.
This allows to obtain the observed frame templates for redshift $z_j$, sampled at the same wavelengths as F($\lambda$), by just shifting one position the indices of the templates for $z_{j-1}$, with no need for a new resampling or interpolation.

For every redshift $z_j$ and template $k$, the routine performs a least squares fit in which the template flux is scaled to fit the spectrum. Since the overall continuum slope provides little information on the redshift of the source, the flexibility of the fit is increased by allowing for a wavelength-dependent scaling factor. That is, the spectrum F($\lambda$) is fitted to a function of the form:
\begin{equation}
f_k(\lambda,z_j) = (a_k(z_j) + b_k(z_j) \cdot \log \lambda ) \cdot S_k\Big{(}\frac{\lambda}{1+z_j}\Big{)}
\end{equation}
where $a$, $b$ are free parameters.
 
This flexible scaling helps the templates obtain better fits even if their continuum slope is somewhat different from that of the spectrum. As a consequence the fit becomes more sensitive to small scale features in the spectrum.

In a standard template fitting, the goodness of fit is quantified by the reduced $\chi^2$ statistics:
\begin{equation}
\chi^2_k(z_j) = \frac{1}{N_{jk}-2} \sum _i^{N_{jk}} \Bigg{(}\frac{F(\lambda_i) - f_k(\lambda_i,z_j)}{\sigma_i}\Bigg{)}^2 
\end{equation}
where $N_{jk}$ is the number of $\lambda$ values in which the (resampled) spectrum and the redshifted template $k$ overlap
at redshift $z_j$, and $\sigma_i$ is the one sigma uncertainty in $F(\lambda_i)$.

The likelihood of a given redshift and template pair ($z$, $T$) is then $\mathscr{L}_T(z) \propto e^{-\chi^2_T(z)}$, and assuming all templates and redshifts have the same probability \emph{a priori}, the maximum likelihood (ML) solution is simply the ($z$, $T$) pair that maximizes $\mathscr{L}_T(z)$.

The ML solution also assumes that the template set is exhaustive (includes all possible types), but this condition is difficult to meet with samples of MIR spectra because of the high number of physical processes involved. 

When none of the templates is an accurate model of the spectrum, the $\chi^2$ minimisation favours those templates that best reproduce the overall shape of the continuum even if small scale features are poorly fit, simply because the latter only affect a small fraction of the $\lambda_i$.
Such behaviour is unwelcome, since the narrow spectral features are crucial to obtain an accurate redshift estimate, while the continuum curvature and slope changes only provide rough redshift indications.

To overcome this limitation, we propose a new algorithm for finding the most probable redshift value, called ``maximum combined pseudo-likelihood'' (MCPL).
 
The main features of MCPL are: a) it searches for local --instead of absolute-- maxima in $\mathscr{L}_T(z)$; b)
it combines information on the local maxima found by all templates to produce a pseudo-likelihood distribution as a function of redshift.

The rationale behind this approach is that the broadband SED of the source determines the general shape of $\mathscr{L}_T(z)$, while narrow spectral features cause high frequency variations in $\mathscr{L}_T(z)$ as they correlate (or not) with features in the template. A good correlation of several features at a certain redshift produces a sharp peak in $\mathscr{L}_T(z)$ that signals a candidate redshift solution.
The spurious alignment of a few features or noise spikes in the spectrum and the template can also produce a peak in $\mathscr{L}_T(z)$ at a wrong redshift. This peak can even be higher than the $\mathscr{L}_T(z)$ value at the actual redshift if the template is a poor model for the spectrum. However, such chance alignments tend to appear at different redshift for each template, while the legitimate peak always occurs at the same (actual) redshift. Because of this, combining information on the position and strength of local maxima produced by all the templates has the potential to yield a more robust redshift estimate compared to considering only the best fitting template.

This idea is implemented by a ``filter'' function that zeroes all values of $\mathscr{L}_T(z)$ except those corresponding to local maxima:

\begin{equation}
\mathscr{L}^*_T(z) = 
\begin{cases} \mathscr{L}_T(z) & \text{if local maximum}\\
0 & \text{otherwise}
\end{cases}
\end{equation}

The combined, filtered likelihood distribution is then the sum over all the templates: 

\begin{equation}
\mathscr{L}^*(z) = \sum_i^{N_T} \mathscr{L}^*_i(z)
\end{equation}

The filtering implies that each template promotes only those redshift values at which it finds a (partial) correspondence of features with the spectrum.

The information provided by the continuum is not lost, though, since the height of the local peaks in $\mathscr{L}_T(z)$ still indicates the goodness of fit between spectrum and template at those particular redshifts.

Since the template set cannot reproduce all the possible combinations of spectral features (that is, it is not complete), even the best fitting template at the correct redshift is not in general an accurate model for the intrinsic spectrum of the source. In other words, the differences between spectrum and template cannot be accounted for by flux uncertainties alone. Therefore, the reduced $\chi^2$ increases with increasing SNR, and produces $\chi^2$ $\gg$ 1 even for fits that a visual inspector would consider very good. 

The exponential dependency of $\mathscr{L}$ with $\chi^2$ implies that the peak with lowest $\chi^2$ usually dominates the resulting combined $\mathscr{L}^*(z)$, making the contribution from other peaks negligible and turning the MCPL algorithm into simple ML.

To overcome this undesired effect, the likelihood function $\mathscr{L}$ can be substituted with a ``pseudo-likelihood'' $q$ that is proportional to the inverse of the reduced chi-squared statistics:
\begin{equation}
 q_T(z) = \frac{a(z,T,\theta)}{\chi^2_T(z)}
\end{equation} 
where the \emph{prior} $a(z,T,\theta)$ captures all the information about the source to be weighted in the selection of the best redshift estimate, such as the observed flux density in a given band, or the \emph{a priori} probability of any ($z$, $T$) combination.

Using $q_T(z) \propto 1/\chi^2_T(z)$ instead of $\mathscr{L}_T(z) \propto e^{-\chi^2_T(z)}$ still favours the lowest value of $\chi^2$, but lets other local peaks have some influence on the final solution.

The ``combined pseudo-likelihood'' function $Q(z)$ is then defined as:
\begin{equation}
Q(z) = \sum_i^{N_T} q^*_i(z)
\end{equation}
where $q^*_i(z)$ is the filtered $q_T(z)$ for template $i$ in which all values other than local maxima have been zeroed.

Variations among templates in the profile of a resolved spectral feature like the 7.7 \um PAH complex or the $\sim$10 \um silicate feature cause that different templates produce peaks at slightly different redshifts. This results in tight clusters of nearby peaks in the combined $Q(z)$. If the redshift difference between the peaks in a cluster is comparable to the theoretical redshift resolution the spectrum is capable of, it can be assumed that all these peaks represent the same redshift solution, although with some dispersion.

To compensate for this, $Q(z)$ is convolved with a gaussian kernel (whose full width at half maximum is twice the redshift resolution) to produce a smoothed $Q_s(z)$. The final solution is then the redshift $\zbest$ that maximizes $Q_s(z)$.

In regular SED-fitting ML photometric redshifts, error bars for the redshift estimate can be computed using the $\Delta\chi^2$ method \citep[e.g.][]{Anvi76,Bolzonella00}. This method assumes the probability distribution for the minimum of $\chi^2(z)$ ($\chi^2_{\rm{min}}$) is the $\chi^2$ distribution for $n$ degrees of freedom \citep{Press92}.
Nevertheless, even for broadband photometric redshifts the $\chi^2$ distribution is not a realistic description of the actual redshift uncertainties, because the model is not linear in the fitting parameters (namely, the redshift) and there are degeneracies between redshift and galaxy SED types \citep{Oyaizu08}.

Like $\chi^2_{\rm{min}}$, the peak value of $Q_s(z)$ depends mainly on the SNR of the spectrum, and does not provide a direct estimation of the reliability of the redshift solution. Still, our results with MIR spectra indicate that the value of $Q_s(\zbestu)$, if normalised to the integral over the entire range of redshifts:

\begin{equation}
\gamma = Q_s(\zbestu)/I, \hspace{1cm} I = \int_{zmin}^{zmax} Q_s(z') dz'
\end{equation}

provides valuable information regarding the strength of the redshift solution. A value of $\gamma$ close to 1 indicates that the peak at $\zbest$ clearly dominates the $Q_s(z)$ distribution, and thus the redshift estimate should be reliable. Conversely, a very low $\gamma$ indicates that there are many secondary peaks with similar strength, and the redshift estimate is unreliable.

Another related parameter, useful to estimate the degree of degeneracy, is the ratio $R$ between the $Q_s(z)$ values for the highest and second-highest peaks. A ratio close to 1:1 indicates peaks of comparable strength and reveals a significant probability of catastrophic redshift error. 

\section{Sample selection}

The spectra used here were selected from the \textit{Spitzer}/IRS ATLAS project (ATLAS-IRS, Hern\'an-Caballero \& Hatziminaoglou 2011), which compiles MIR spectra and ancillary data from 739 extragalactic sources at 0 $<$ $z$ $<$ 3.7.

The parent sample is composed of all the ATLAS-IRS sources with a known spectroscopic redshift from optical or near-infrared spectroscopy ($\zspecu$). To ensure enough spectral coverage, 20 sources observed in only one of the four IRS modules were discarded. 11 additional sources where selected as templates (see \S\ref{sectiontemplates}) and removed from the sample to avoid circularity issues. 

The information content in each spectrum was computed using the net significance ($\mathscr{N}$), defined as the maximum cumulative SNR of the spectrum \citep{Pirzkal04}.
The 3 sources with lowest net significance values ($\mathscr{N}$ $<$ 100, corresponding to median SNR per pixel $\lesssim$ 0.8) were also removed from the sample.

\begin{figure} 
\begin{center}\hspace{-0.3cm}
\includegraphics[width=8.5cm]{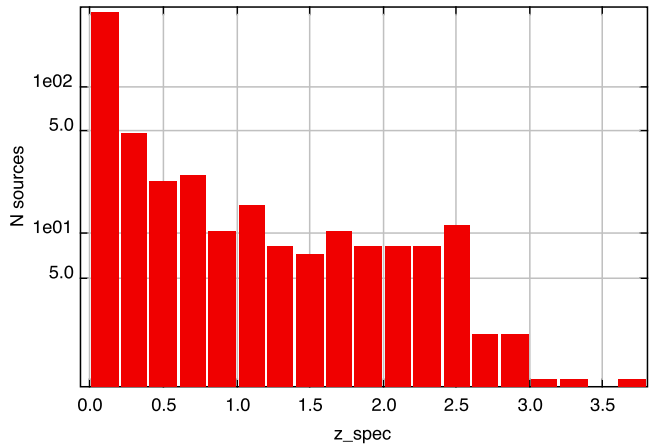}
\end{center}
\caption[]{Redshift distribution for the 491 sources in the main sample.\label{zspec-histog}}
\end{figure}

The remaining 491 sources constitute the main sample of this work. Their redshift distribution is shown in Figure \ref{zspec-histog}. 

About half of the sample is at redshift $\zspec$ $<$ 0.15, while only $\sim$20\% is at $\zspec$ $>$ 1. This distribution is similar to that of the entire set of extragalactic sources with \textit{Spitzer}/IRS spectroscopy and known spectroscopic redshifts \citep{Lebouteiller11}. 

321 sources (65\% of the sample) are classified as AGN in the optical, including 124 optical quasars (QSO), 46 obscured quasars (QSO2), 12 type 1 Seyferts (Sy1), 73 intermediate type Seyferts (Sy1.X), 60 type 2 Seyferts (Sy2), and 56 LINERs.

In the MIR, 285 sources (58\% of the sample) are classified as AGN-dominated, while 181 are starburst-dominated and 10 are classified as composites with roughly equal contributions from the AGN and starburst to the infrared output of the galaxy. See \citet{Hernan-Caballero11} for further details on the optical and MIR classifications.

\section{Templates}\label{sectiontemplates}

\begin{figure} 
\begin{center}\hspace{-0.3cm}
\includegraphics[width=8.5cm]{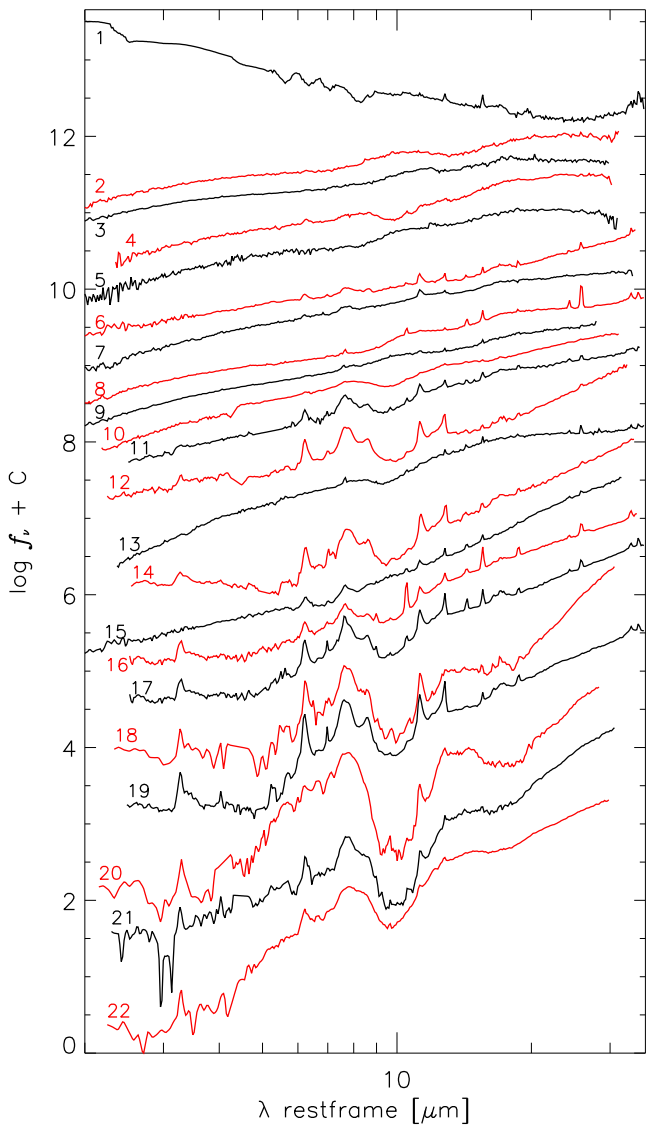}
\end{center}
\caption[]{The 22 spectral templates used by the SED-fitting routine, ordered by increasingly red continuum. Numbers correspond
to row indices in table \ref{table-templates}. Alternative red and black colours are used for clarity only.\label{templates}}
\end{figure}

To obtain reliable redshift estimates, it is essential that the templates cover as much as possible the restframe spectrum of the source at any redshift, since insufficient overlap between spectrum and template increases the probability of obtaining a good fit at a wrong redshift due to chance alignments of spectral features.

For spectra observed in the four IRS channels (5--35 \uu), a redshift search range 0 $<$ $z$ $<$ 4 implies that the templates should ideally span the entire 1--35 \um restframe range. In practice, shorter wavelength coverage suffices, as long as the template and spectrum share enough overlap in the entire redshift search range. 

A large number of spectral templates were generated based on IRS spectra from ATLAS-IRS and from the Cornell Atlas of Spitzer/IRS Spectra \citep[CASSIS;][]{Lebouteiller11} using several methods:

a) for a sample of local luminous and ultra-luminous infrared galaxies, their AKARI 2.5--5 \um spectrum \citep{Imanishi10} is used to extend their IRS spectrum down to restframe $\sim$ 2 \uu.
b) composite spectra of several samples of $z$ $>$ 1 quasars (selected by their
restframe 3.6 to 5.6 \um continuum slope and/or strength of the silicate feature) are used
to extend the IRS spectrum of individual, lower redshift quasars with good SNR.
c) individual and composite spectra of $z$ $>$ 0.5 radiogalaxies are used to extend individual spectra of low redshift radigalaxies.
d) another template is produced from the composite spectra of radiogalaxies with strong emission in the [SIV] 10.51 \um line.
e) An early-type non-active galaxy template is obtained by extending the IRS spectrum of 
NGC 5011 with the elliptical galaxy template from \citet{Coleman80}.

In addition, archival \textit{ISO}/SWS 2--45 \um spectra of NGC1068, M82, and the Circinus galaxy are also included as templates.

A selection procedure was devised to identify the best performing templates. First, the filtered pseudo-likelihood function, $q_{i,j}(z)$, is calculated for every pair \{$i$,$j$\} of template and spectrum. Then, subsets of templates are given a score based on the number of accurate ($\Delta z/(1+z)$ $<$ 0.01) redshift solutions obtained with that particular subset. A penalisation factor depending on the number of templates is also included to discourage large template sets.

An iterative process finds the subset that maximizes the score by randomly adding or removing one or several templates to the subset, keeping only those changes that increase the score until no further increases are possible. This process is run several times to ensure it always converges to the same template set.

Note that since the MCPL redshift solution for any source depends on the whole template set (and not just the template obtaining the best fit) the optimisation would discard a template that produces good fits for a few unusual sources if it degrades the solution for many others. Nevertheless, it is unlikely that a chance alignment of some features in a template that does not match the overall SED of a source can produce a peak in $q_{i,j}(z)$ strong enough to overshadow the combined peaks produced by the remaining templates at the actual redshift.

The final set, containing 22 templates, is shown in Figure \ref{templates} and listed in Table \ref{table-templates}. The templates derived from spectra of sources in the sample produce, as expected, very good fits for these particular sources. To avoid misrepresenting the actual accuracy of the method, these sources are removed from the sample.
 
\section{Results}

\zcolors was run on the sample of 491 spectra selected from ATLAS-IRS with the template set described in the previous section.

The spectra and templates are resampled to a constant spectral resolution $R$ = 500, which provides a redshift resolution $\frac{\Delta z}{1+z}$ = 0.002. In test runs, higher resolution values increase the computational cost with no significant gain in accuracy of the redshift solutions.  
The search range for redshifts is -0.05 $\leq$ $z$ $\leq$ 4, with the extension to small negative values being important to properly identify the peak of $q_T(z)$ in nearby galaxies ($z$ $\sim$ 0).  

All templates are assumed to have the same \emph{a priori} probability. The only \emph{prior} introduced is a luminosity limit, aimed at preventing bright sources from obtaining high redshift estimates that would imply unrealistically high luminosities.
The luminosity limit is conservatively put at $\nu$L$_{\nu}$ = 5$\times$10$^{47}$ erg s$^{-1}$, which is just above the most luminous source in the sample. 
For every source, the redshift ($z_{\rm{cut}}$) corresponding to this luminosity is found,
and the \emph{prior} $a(z)$ is then defined by:

\begin{equation}
a(z) = 
\begin{cases} 1 & \text{if } z \leq z_{\rm{cut}}\\
0 & \text{if } z > z_{\rm{cut}}
\end{cases}
\end{equation}

The luminosity limit achieves a $\sim$30\% reduction in the number of catastrophic errors ($\Delta z$/(1+$z$) $>$ 0.1) relative to a flat \emph{prior}. This suggests
a more elaborate \emph{prior} including luminosity limits that depend on the SED and observed flux of the source could probably help to further reduce degeneracies. 

Table \ref{results-table} contains the redshift solutions obtained by the ML and MCPL algorithms for all the sources in the sample. 

\subsection{Accuracy of redshift solutions}

\begin{figure} 
\begin{center}\hspace{-0.3cm}
\includegraphics[width=8.5cm]{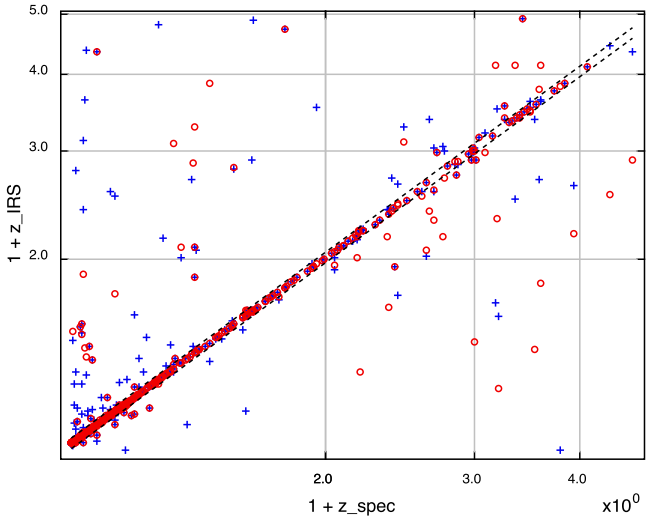}
\end{center}
\caption[]{Comparison between the redshifts derived from template fitting ($\zIRSu$)  and those from optical spectroscopy ($\zspecu$). Plus signs mark solutions from the ML algorithm, while open circles mark those of the MCPL selection algorithm. The dotted lines enclose those sources with redshift accuracy $\delta$ $<$ 0.02.\label{zspec-zIRS}}
\end{figure}

The accuracy of the $\zIRS$ estimates is evaluated by comparing them to the redshifts from 
optical or NIR spectroscopy ($\zspecu$). The error in the $\zIRS$ value is represented by $d$ = $(\zIRS - \zspecu)/(1+\zspecu)$ and its modulus, $\delta$ = $\vert d \vert$, defines the accuracy of the redshift solution.

Figure \ref{zspec-zIRS} shows the correlation between $\zIRS$ and $\zspec$ values for both, the ML and MCPL selection algorithms. 86\% of MCPL solutions and 78\% of ML are enclosed within the parallel lines that represent accuracy $\delta$ $<$ 0.02, but typical accuracies are much higher. 

\begin{figure} 
\begin{center}\hspace{-0.3cm}
\includegraphics[width=8.5cm]{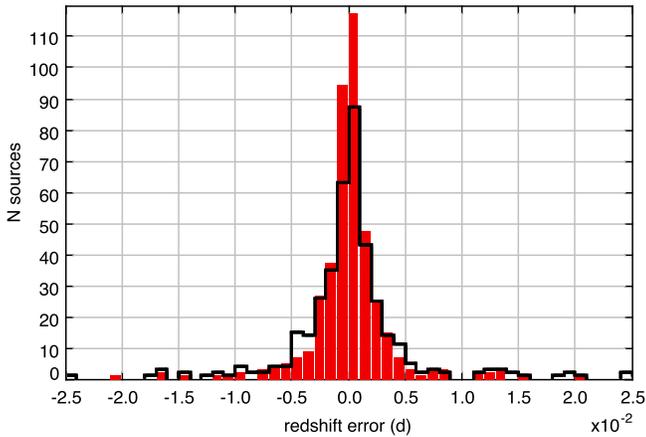}
\end{center}
\caption[]{Distribution of redshift errors ($d$ = $(\zIRS - \zspecu)/(1+\zspecu)$) obtained using the ML (solid line) and MCPL (solid bars) selection algorithms. \label{dz-histog}}
\end{figure}

The number of outliers ($\delta$ $>$ 0.02) is 69 for MCPL and 106 for ML. Excluding them, errors for MCPL (ML) solutions average 2.0$\times$10$^{-5}$ (-4.5$\times$10$^{-5}$) with a standard deviation of 0.0033 (0.0046). 
This indicates there is no significant bias in the 
redshift estimates and the typical errors are just a fraction of the spectral resolution of IRS ($\delta\lambda$/$\lambda$ $\sim$ 0.008--0.016, depending on wavelength). The distribution of redshift errors is approximately gaussian, with full width at half maximum 0.0047 for MCPL and 0.0056 for ML (see Figure \ref{dz-histog}). 

\begin{figure} 
\begin{center}\hspace{-0.3cm}
\includegraphics[width=8.5cm]{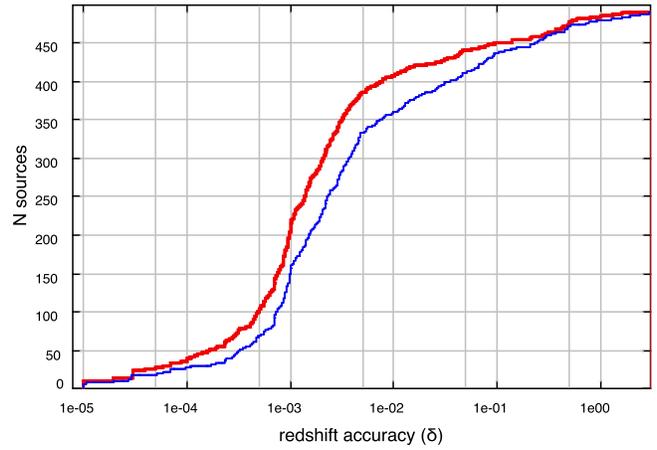}
\end{center}
\caption[]{Cumulative distribution of the accuracy parameter $\delta$ = $\vert \zIRS - \zspec \vert$/(1+$\zspec$).The thin solid line represents results for the ML selection algorithm while the thick solid line corresponds to those of the MCPL algorithm.\label{delta-counts}}
\end{figure}

The cumulative distribution of $\delta$ for MCPL and ML solutions is presented in Figure \ref{delta-counts}. Both distributions show similar trends, with a rapid growth in the number of sources up to $\delta$ $\sim$ 0.005 and much slower growth at higher values. The curve for MCPL is consistently over ML in the entire $\delta$ range, but the separation is larger at $\delta$ $\sim$ 0.005. 
The number of sources in the range 0.02 $<$ $\delta$ $<$ 0.1 is 52 for ML versus 29 for MCPL, indicating a higher prevalence of low accuracy solutions in ML. Catastrophic redshift errors ($\delta$ $>$ 0.1) are obtained in 40 and 54 sources for MCPL and ML, respectively. 

Comparison of ML and MCPL solutions for individual sources reveals that in 2/3 of the sample (318 sources) $\delta$ values from both algorithms are within 10\% of each other. In another 132 sources the MCPL solution is clearly more accurate (sometimes by several orders of magnitude), while only in 37 cases is the ML significantly better.

The accuracy advantage of MCPL over ML is clearer at low redshift: at $z$ $<$ 0.5
MCPL outperforms ML in 96 cases versus 16 for ML, while at $z$ $>$ 1 they are levelled, with each of them wining in 19 cases. 
This is probably a consequence of the decrease in the average SNR with redshift. A lower SNR reduces the contrast of the legitimate peak in $Q_s(z)$ and makes it easier for spurious peaks to obtain comparable strength, increasing the risk of degeneracies. 
   
\begin{figure} 
\begin{center}\hspace{-0.3cm}
\includegraphics[width=8.5cm]{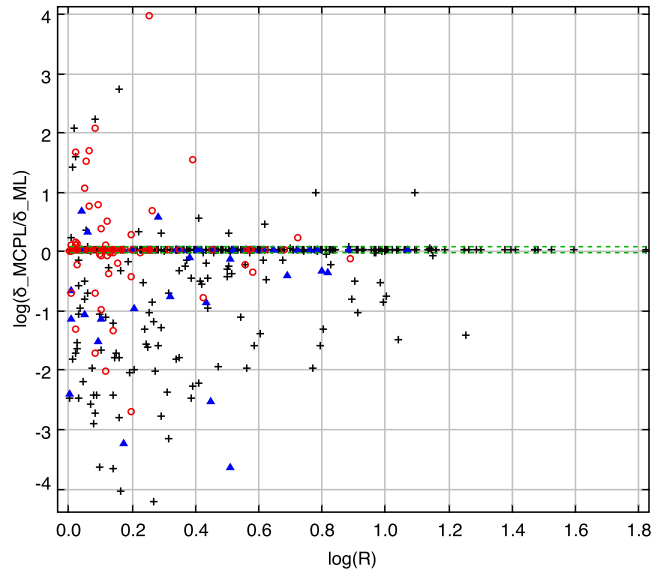}
\end{center}
\caption[]{Logarithm of the ratio between accuracies of the MCPL and ML redshift solution versus degeneracy parameter (R) of the MCPL solution for sources at $\zspec$ $<$ 0.5 (plus signs), 0.5 $<$ $\zspec$ $<$ 1 (solid triangles), and $\zspec$ $>$ 1 (open circles). Negative values indicate a higher accuracy for the MCPL solution compared to ML. The dotted lines enclose the region corresponding to MCPL and ML solutions within 10\% of each other, which contains 65\% of the sources in the sample.\label{delta-ratio}}
\end{figure}

Figure \ref{delta-ratio} represents the ratio between $\delta$ values of the MCPL and ML
solutions versus the ``degeneracy parameter'' (R), defined as the ratio between the highest and second highest peaks in $Q_s(z)$. Most sources with ML solutions significantly more accurate than MCPL have $\log$(R) $<$ 0.15, indicating an extreme degree of degeneracy in $Q_s(z)$. In these sources, the two highest peaks in $Q_s(z)$ have very similar strength, and it is thus no surprise that MCPL chooses sometimes the wrong one.
Nevertheless, even with a very degenerate $Q_s(z)$ MCPL offers higher reliability than ML: 
from 138 sources with $\log$(R) $<$ 0.15, the MCPL solution is accurate ($\delta$ $<$ 0.02) while ML is an outlier ($\delta$ $>$ 0.02) in 26 cases, versus only 7 the other way around. In another 63 cases both are accurate and in the remaining 40 both are outliers.
 
\subsection{Reliability of individual solutions}\label{reliability-subsection}

\begin{figure} 
\begin{center}\hspace{-0.3cm}
\includegraphics[width=8.5cm]{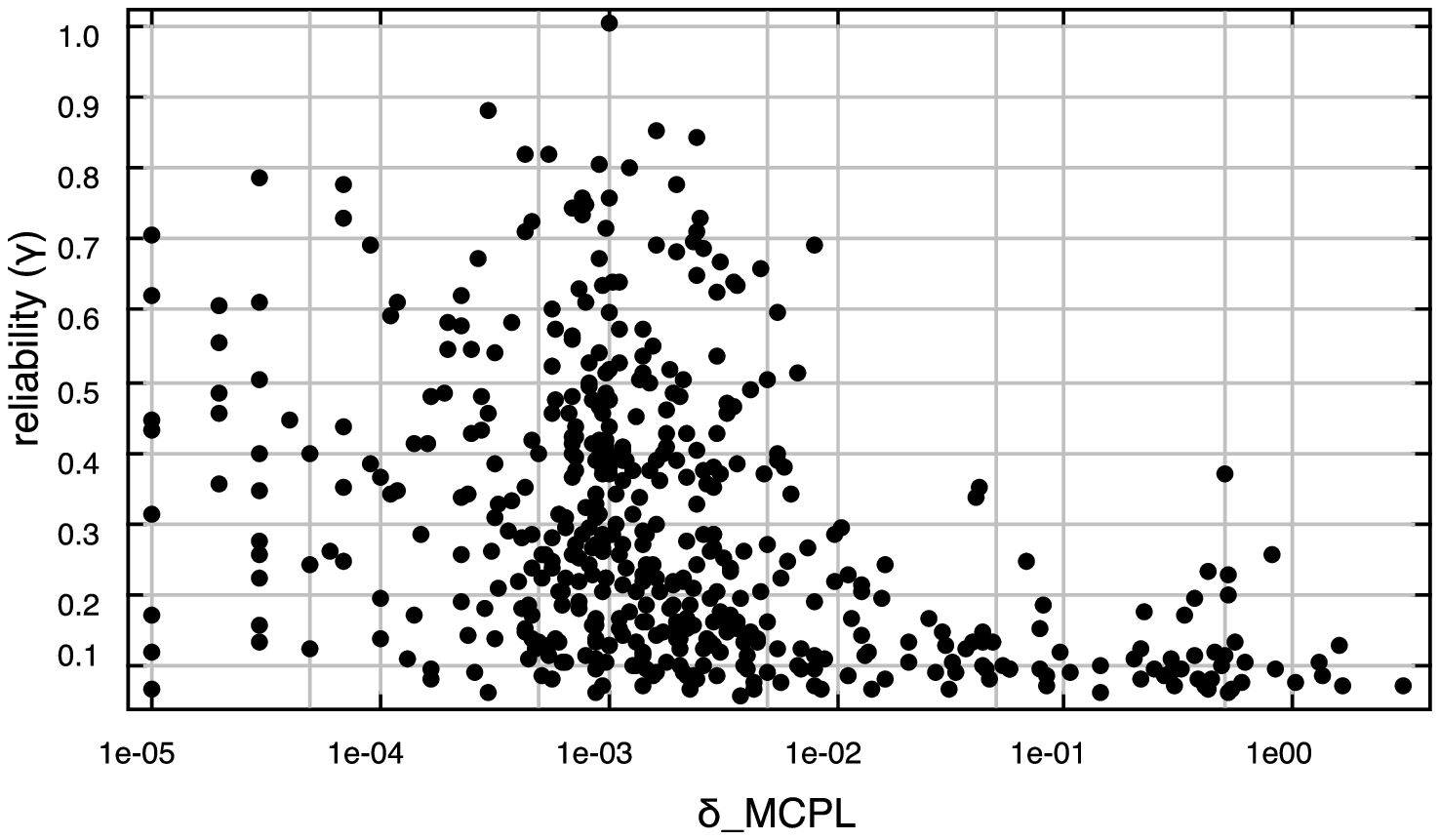}
\includegraphics[width=8.5cm]{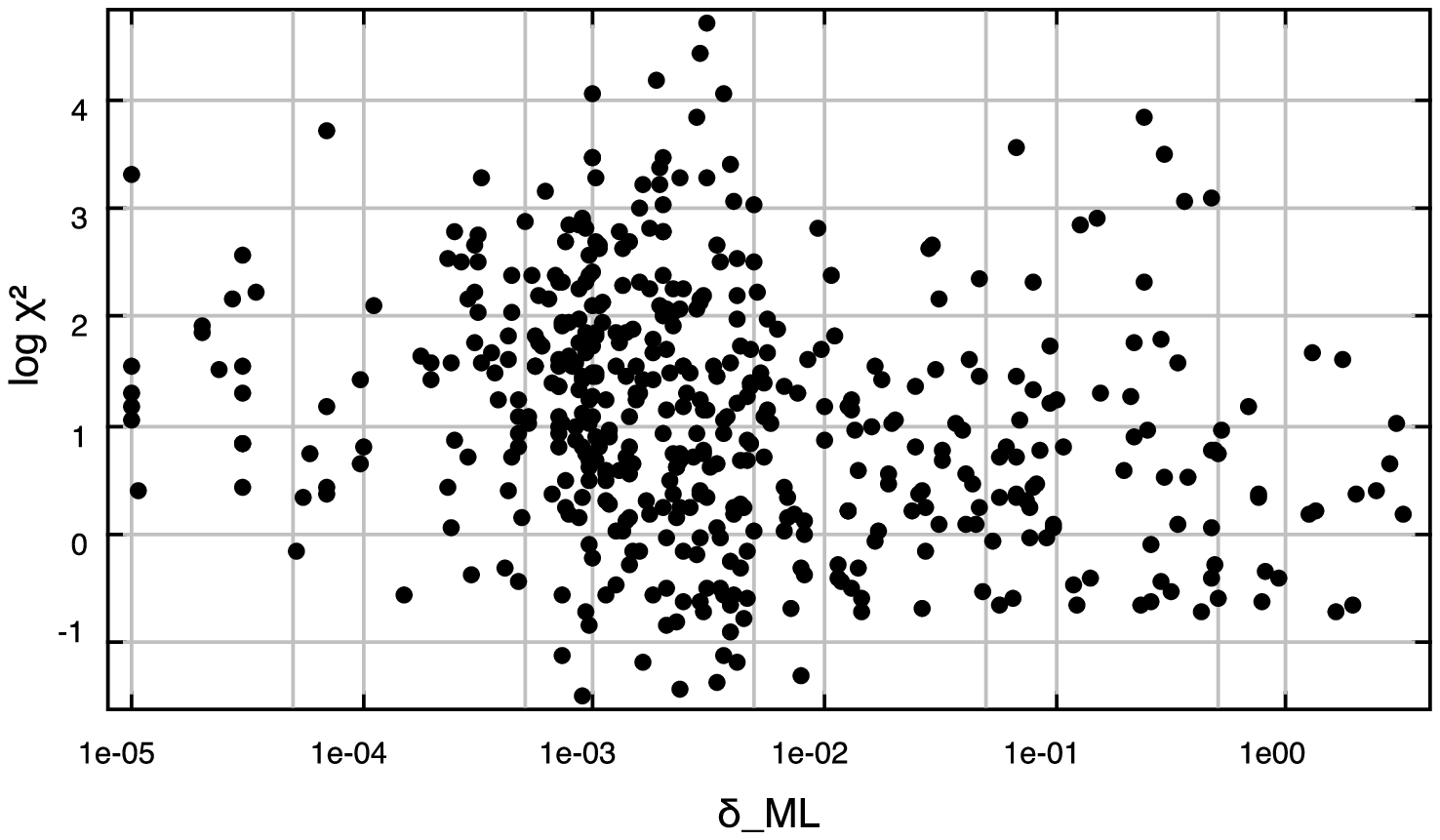}
\end{center}
\caption[]{Top: relation between reliability ($\gamma$) and accuracy ($\delta$) parameters for MCPL solutions. Bottom: logarithm of the reduced $\chi^2$ statistic versus $\delta$ for ML solutions.\label{gamma-chi2}}
\end{figure}

Apart from obtaining a high rate of accurate solutions, it is important to know the reliability of individual solutions. In \S\ref{themethod} we anticipated that the $\gamma$ parameter can provide such information for the MCPL algorithm. 

The upper panel in Figure \ref{gamma-chi2} represents $\gamma$ versus $\delta$ for the MCPL solutions of the 491 sources in the sample.
Accurate solutions obtain $\gamma$ values spanning the whole $\sim$0.05--1 range, while outliers concentrate at $\gamma$ $\lesssim$ 0.15, with few cases above that value. 

Table \ref{reliability-table} demonstrates that the reliability of MCPL solutions increases monotonically with $\gamma$, both in terms of the dispersion and median of $\delta$ values and in the frequency of outliers. 
This implies that by setting an appropriate threshold value for $\gamma$ and selecting only the sources that surpass it, it is possible to obtain subsamples of sources with very reliable redshift estimates, albeit at a cost in completeness.

In contrast, $\chi^2$ values for ML solutions (lower panel in Figure \ref{gamma-chi2}) do not show an increase with $\delta$. On the contrary, the higher $\chi^2$ values occur preferably in sources with accurate redshifts, because these are usually the ones with higher SNR spectra. In other words: the $\chi^2$ statistic correlates with the SNR of the spectrum, because at high SNR differences in the profile and strength of spectral features between spectrum and template are evident, while a very noisy spectrum blurs its features to the point that some of the templates can be considered an accurate model even at the wrong redshift. Therefore, the value of the absolute minimum in $\chi^2(z)$ cannot be used to identify the reliable ML solutions. 

Since the MCPL algorithm offers higher redshift accuracy with a lower number of outliers, and also provides an indication on the reliability of the redshift solution, it can be considered superior to ML for this purpose. In the remaining sections, only results from MCPL will be discussed.

\subsection{Selection efficiency and completeness}

Let $N_g(D)$ be the number of sources with accuracy $\delta$ $<$ $D$, $N_s(G)$ the number of sources with $\gamma$ $>$ $G$, and $N_{sg}(D,G)$ the number of sources  with $\delta$ $<$ $D$ and $\gamma$ $>$ $G$.
The selection efficiency ($\epsilon$) and the completeness ($\kappa$) are then defined by:
\begin{equation}
\epsilon(G,D) = \frac{N_{sg}(D,G)}{N_s(G)},\hspace{1cm} \kappa(G,D) = \frac{N_{sg}(D,G)}{N_g(D)}
\end{equation}
  
\begin{figure} 
\begin{center}\hspace{-0.3cm}
\includegraphics[width=8.5cm]{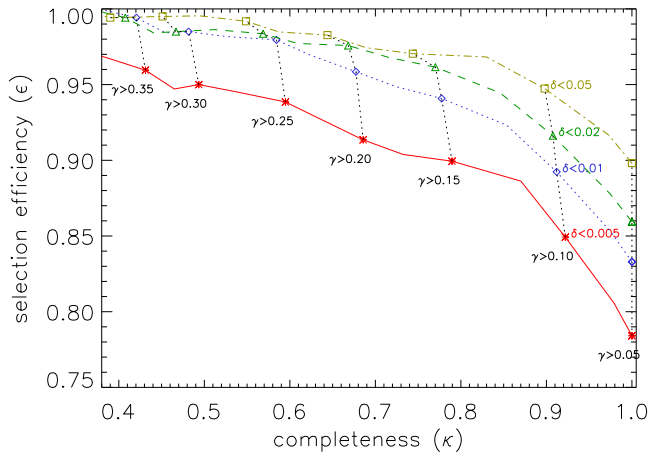}
\end{center}
\caption[]{Selection efficiency versus completeness as a function of the threshold value used for the reliability parameter $\gamma$ for redshift solutions with accuracy $\delta$ $<$ 0.005 (solid line), $\delta$ $<$ 0.01 (doted line), $\delta$ $<$ 0.02 (dashed line) and $\delta$ $<$ 0.05 (dot-dashed line).\label{epsilon-kappa}}
\end{figure}

As usual, there is a trade-off between completeness and selection efficiency, with either of them increasing only at the expense of the other. 
The relationship between $\epsilon$ and $\kappa$ for a grid of values of
$D$ and $G$ is shown in Figure \ref{epsilon-kappa}.

For $D$ values in the range 0.005--0.05, the completeness has a very small (but consistent) dependency on $D$ that reflects the reduction in the average $\gamma$ for higher values of $\delta$. The dependency with $G$ is much stronger due to the large number of sources with low $\gamma$ values.

Efficiencies for $D$ = 0.01, 0.02 and 0.05 converge rapidly to $\epsilon$ = 1 with increasing $\gamma$, 
because there are almost no sources with high $\gamma$ values and $\delta$ $>$ 0.01.
Nevertheless, for $D$ = 0.005 there is slower growth, since some sources with very high $\gamma$ values have accuracies $0.005$ $<$ $\delta$ $<$ 0.01.

\subsection{Dependency with the MIR SED}\label{dependency-sed}

Differences in the MIR SED of starburst-dominated versus AGN-dominated sources cause important variations in the average accuracy of the redshift solutions depending on the MIR SED type.

Normal and starburst galaxies usually have very prominent PAH bands that are easily identified even in low SNR spectra, and almost always obtain very accurate redshifts. On the other hand, sources dominated by AGN emission usually show a flat continuum with no high contrast features, and the redshift determination relies on
broad and shallow silicate features (in emission or absorption) and/or unresolved emission
lines, making it much harder to distinguish the peak in $Q_s(z)$ corresponding to the actual redshift of the source.

In \citet{Hernan-Caballero11} the ratio $r_{\rm{PDR}}$ of total PAH luminosity to the integrated restframe 5--15 \um luminosity is used to classify the ATLAS-IRS spectra into starburst-dominated (MIR\_SB, $r_{\rm{PDR}}$ $>$ 0.15) and AGN-dominated (MIR\_AGN, $r_{\rm{PDR}}$ $<$ 0.15). This is roughly equivalent to imposing a threshold EW $\sim$0.2 \um in the equivalent width of the 6.2 \um or 11.3 \um PAH bands, which corresponds to roughly equal contributions from the starburst and AGN to the infrared luminosity of the galaxy \citep{Hernan-Caballero09}.
MIR\_AGN sources are further separated into silicate emission (MIR\_AGN1) and silicate absorption (MIR\_AGN2) sources. 
Table \ref{MIRtype-table} summarizes the accuracy and reliability statistics for these populations.

Sources dominated by star formation in their MIR spectra (MIR\_SB) almost always obtain accurate redshifts. Among the 182 MIR\_SB galaxies in the sample, MCPL solutions include only 2 outliers ($\delta$ $>$ 0.02), namely Murphy19 and NGC 4579.
The optical redshift of the submillimeter galaxy Murphy19 (SDSS J123716.59+621643.9) is $\zspec$ = 0.557 \citep{Wirth04}, but \citet{Murphy09} give $z$ = 1.82 based on the IRS spectrum. Although our solutions are consistent with the later ($z_{\rm{MCPL}}$ = 1.806; $z_{\rm{ML}}$ = 1.795), the second highest peak in $Q_s(z)$ is at $z$ = 0.5495, indicating the optical redshift is confirmed with $\gtrsim$98\% confidence (see \S\ref{redshift-degeneracies}).
NGC 4579 (M 58) is a local spiral galaxy. The IRS spectrum contains emission from the LINER nucleus and its surroundings, and shows very intense H$_2$ lines combined with an unusual PAH spectrum (bright 11.3 \um PAH emission but very weak 6.2 and 7.7 \um bands). The lack of templates with significant H$_2$ emission is probably the cause of the wrong redshift solution for this source. In spite of that, the second highest peak in $Q_s(z)$ coincides with the optical redshift.

\begin{figure} 
\begin{center}\hspace{-0.3cm}
\includegraphics[width=8.5cm]{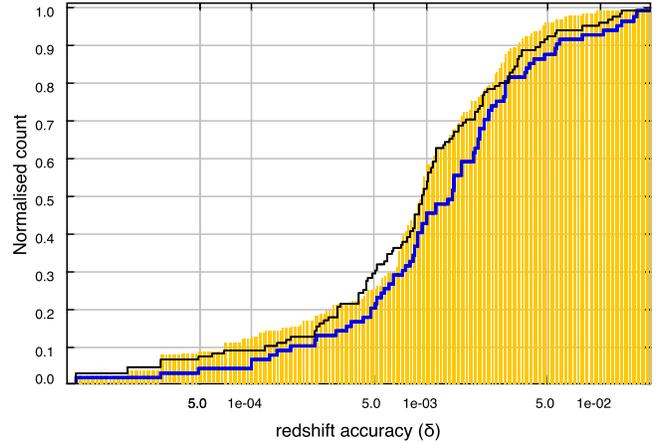}
\end{center}
\caption[]{Normalized, cumulative distribution of $\delta$ values for MIR\_SB (shaded area), MIR\_AGN1 (thick solid line) and MIR\_AGN2 (thin solid line) sources excluding outliers with $\delta$ $\>$ 0.02.\label{delta-counts-MIRtype}}
\end{figure}

Sources classified as MIR\_AGN have redshifts that are much less reliable compared to those for MIR\_SB. The overall outlier rate is 22\%, but there are strong variations in reliability among MIR\_AGN subclasses: the fraction of outliers is 28\% for MIR\_AGN1 versus 8\% for MIR\_AGN2, and up to 40\% for the MIR\_AGN with no clear silicate emission or absorption.
Nevertheless, if the outliers are removed, the accuracies for the remaining sources show very similar distributions in the MIR\_SB, MIR\_AGN1 and MIR\_AGN2 subsamples (Figure \ref{delta-counts-MIRtype}).
  
\begin{figure} 
\begin{center}\hspace{-0.3cm}
\includegraphics[width=8.5cm]{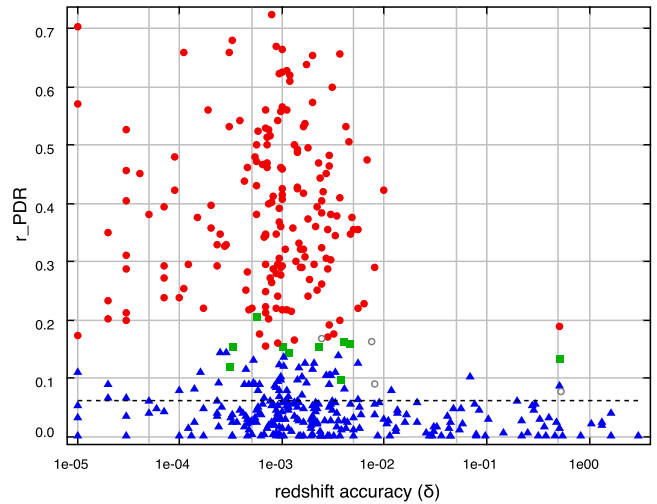}
\end{center}
\caption[]{Ratio of total PAH luminosity to integrated 5--15 \um luminosity ($r_{\rm{PDR}}$) versus redshift accuracy of the MCPL solution for sources classified as MIR\_SB (solid circles), MIR\_AGN (triangles), composite sources (squares), and sources with no MIR classification (open circles). The dashed line marks $r_{\rm{PDR}}$ = 0.06.\label{rPDR-delta}}
\end{figure}

Further insight into the importance of the PAH bands for the accuracy (or lack thereof) of the redshift solution can be obtained from Figure \ref{rPDR-delta}, which shows the strength of the PAH features, represented by $r_{\rm{PDR}}$, versus $\delta$. 

The subsample with $r_{\rm{PDR}}$ $>$ 0.06 includes by definition all the MIR\_SB 
sources ($r_{\rm{PDR}}$ $>$ 0.15), the MIR composites ($r_{\rm{PDR}}$ $\sim$ 0.15), as well as some MIR\_AGN with significant PAH emission. It comprises half of the total sample (242 sources) and has a 2\% rate of outliers and a median accuracy $\tilde{\delta}$ = 9.6$\times$10$^{-4}$.

This demonstrates that detectable PAH emission is sufficient to obtain very accurate and 
reliable redshift estimates, even in high redshift sources with poor SNR spectra. But PAHs are not the only feature capable of providing an accurate estimate, since there are many accurate redshifts down to $r_{\rm{PDR}}$ $\sim$ 0.

\begin{figure} 
\begin{center}\hspace{-0.3cm}
\includegraphics[width=8.5cm]{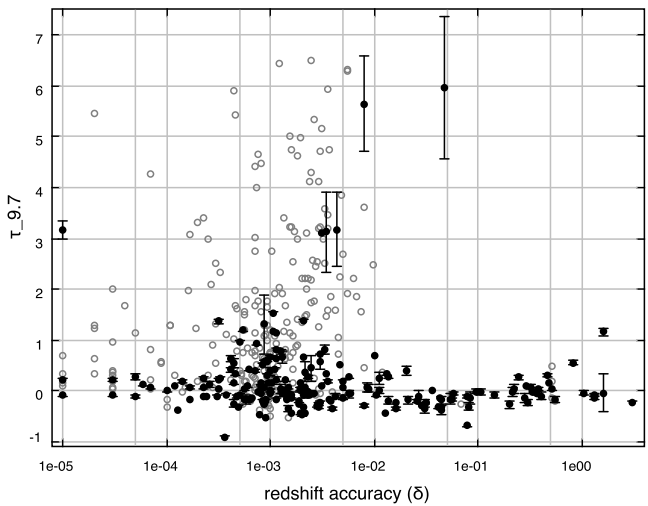}
\end{center}
\caption[]{Optical depth at 9.7 \um (\tausilu) versus redshift accuracy. Solid symbols represent sources classified as MIR\_AGN with very weak or absent PAH bands ($r_{\rm{PDR}}$ $<$ 0.06) while open symbols represent the rest of the sample. Negative (positive) \tausil values indicate silicate emission (absorption).\label{tau97-delta}}
\end{figure}

Figure \ref{tau97-delta} shows the 9.7 \um apparent optical depth (a measurement of the strength of the silicate feature, see \citet{Hernan-Caballero11} for a discussion) versus $\delta$ for the sources with weak or undetected PAH bands ($r_{\rm{PDR}}$ $<$ 0.06).
In this subsample, sources with silicate absorption (\tausil $>$ 0) are much more likely
to obtain accurate redshifts than those with silicate emission (90\% versus 70\%, respectively, with $\delta$ $<$ 0.02), in spite of both populations having similar distributions of $r_{\rm{PDR}}$. This suggests that the silicate feature plays an important role in the redshift determination of sources with weak or no PAH emission. 
The diversity of shapes and lower contrast that the silicate feature presents when it appears in emission might be at least in part responsible for the decreased efficiency in these sources.

\begin{figure} 
\begin{center}\hspace{-0.3cm}
\includegraphics[width=4.42cm]{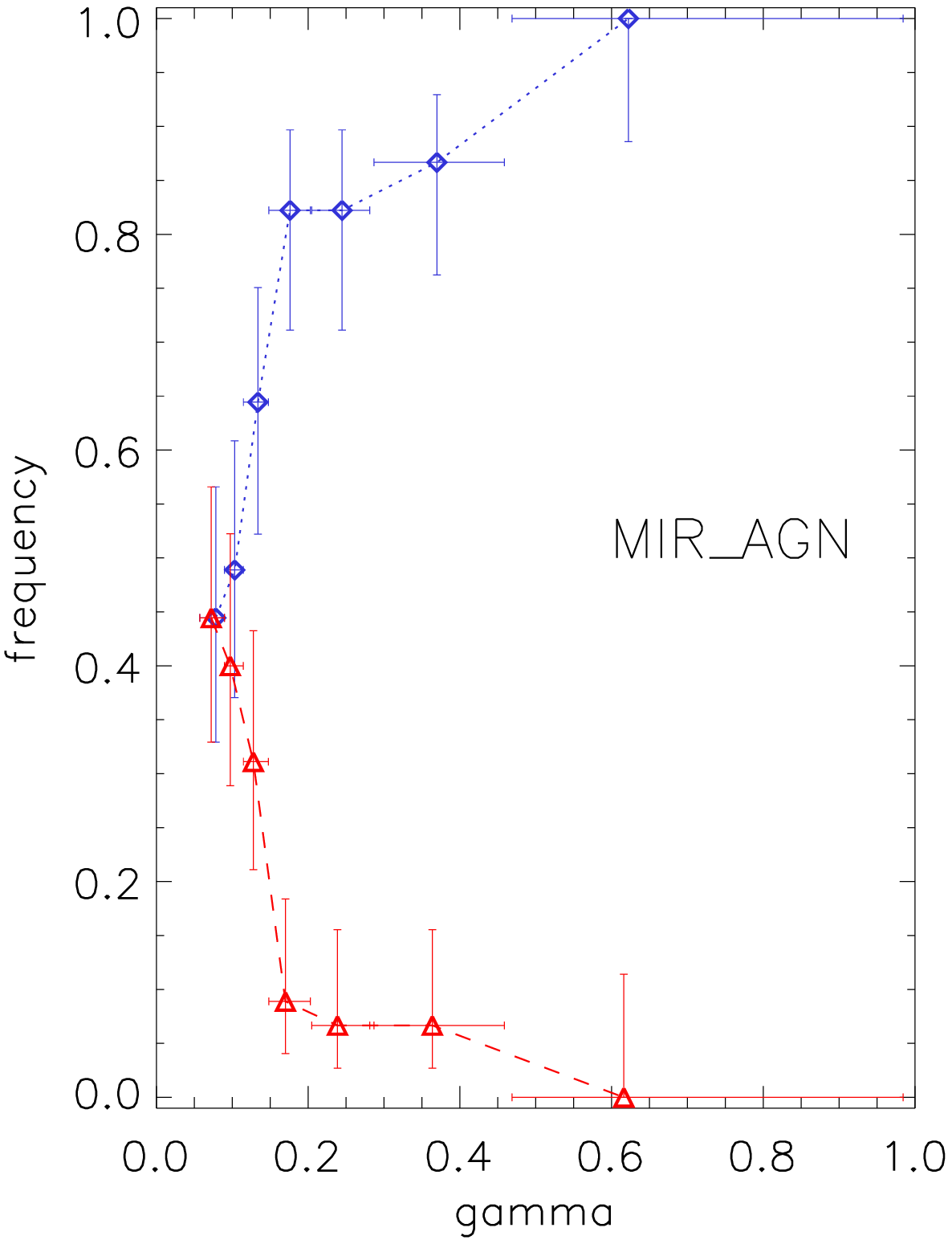}
\hfill
\includegraphics[width=4.08cm]{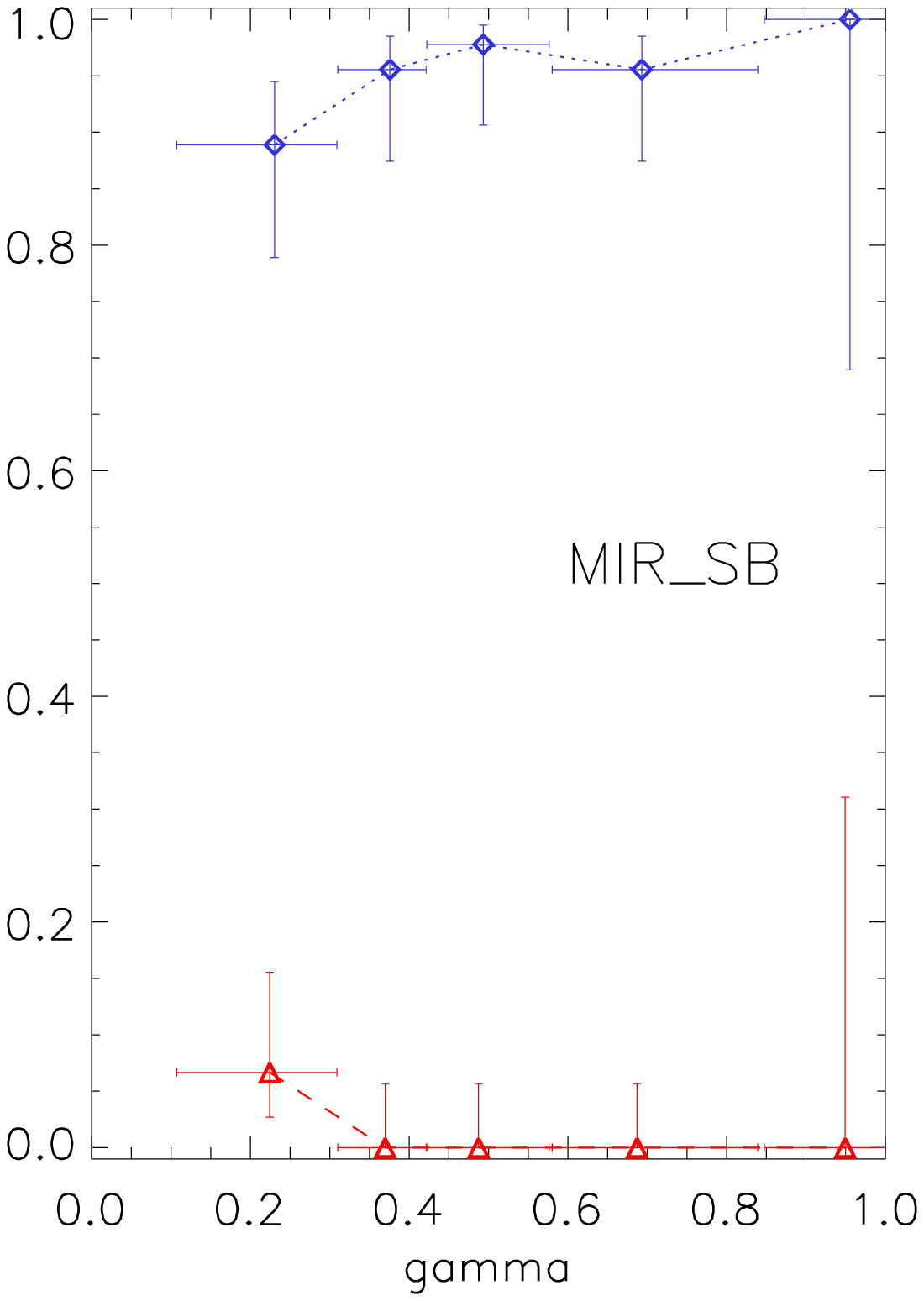}
\end{center}
\caption[]{Frequency of highly accurate solutions ($\delta$ $<$ 0.005, blue diamonds) and outliers ($\delta$ $>$ 0.02, red triangles) as a function of $\gamma$ for the sources classified as MIR\_AGN (left panel) and MIR\_SB (right panel). Each point represents a bin of $\gamma$ containing 45 sources (except for the rightmost which contains the remainder). Horizontal bars represent the $\gamma$ range covered by each bin, while vertical error bars represent the 90\% confidence intervals calculated using the Wilson score formula for binomial distributions.\label{frequencies-good-fail}}
\end{figure}

Albeit the rate of outliers is much higher in MIR\_AGN compared to MIR\_SB sources, the
reliability of redshift estimates within a given $\gamma$ interval is largely independent
on the MIR classification. Figure \ref{frequencies-good-fail} shows the frequencies of
highly accurate solutions ($\delta$ $<$ 0.005) and outliers ($\delta$ $>$ 0.02) as a function of $\gamma$ for the MIR\_AGN and MIR\_SB populations separately. They are found to agree within their 90\% confidence limits.
 
These confidence intervals can be used to put a lower limit on the probability of the redshift solution for a given source having accuracy better than some predefined value, or an upper limit on the probability of being an outlier.
A more detailed model of such probabilities based on a much larger sample of MIR spectra from the CASSIS database is under development, and will be presented elsewhere (Hern\'an-Caballero et al., in preparation).

\subsection{Redshift degeneracies}\label{redshift-degeneracies}

There are 69 outliers ($\delta$ $>$ 0.02) in the sample, 40 of them with catastrophic errors ($\delta$ $>$ 0.1) in their redshift estimates.

Typical values of $\gamma$ for the outliers are low, with 60\% of them below 0.15 (compared to just 20\% in the whole sample). They show multiple secondary peaks in the $Q_s(z)$ distribution, with highest to second-highest ratios ($R$) in the range 1 $<$ $R$ $<$ 3, and in 90\% of cases verifying $R$ $<$ 2 (compared to only 36\% in the sources with $\delta$ $<$ 0.02). 
This suggests that most of these sources have degeneracy issues, and somehow spurious solutions obtain a $Q_s(z)$ value higher than the peak for the actual redshift of the source.

Another possibility that deserves consideration is a wrong or inaccurate optical redshift. 
One way to rule out an error in the optical redshift is to search for a secondary peak in $Q_s(z)$ that matches the $\zspec$ value. If a significant peak is found very close to it, the optical redshift can be confirmed with high probability. On the other hand, the lack of a nearby secondary peak does
not imply that $\zspec$ is wrong, since spectra with very exotic MIR SEDs, strong artefacts or very low signal to noise ratios could produce a very weak peak at the actual redshift that passes unnoticed.  
 
To find out whether there are secondary solutions backing up the $\zspec$ value for the
69 outliers, a routine finds all the peaks in the $Q_s(z)$ distribution that verify $\gamma$ $>$ 0.01, and sorts them by their $\gamma$ value.
48 outliers show at least one such secondary peak within $\delta$ $<$ 0.02 of the $\zspec$ value. They are all listed in Table \ref{secondary-solutions}.

Since the redshift search range is very wide compared to the typical $\delta$ of these solutions, the probability of them occurring that close to $\zspec$ fortuitously is low. If spurious peaks in $Q_s(z)$ are assumed to be randomly distributed in the
redshift search range, the probability for a spurious solution obtaining accuracy $\delta$
or better is:
\begin{equation}
P = \frac{2 \delta}{\ln(1+z_{max})}
\end{equation}
where $z_{max}$ is the upper limit of the redshift search range for that particular source.

For a $Q_s(z)$ distribution with several spurious peaks, the probability that at least one of the $n$ highest peaks is within $\delta$ of $\zspec$ just by chance is then:

\begin{equation}
P_r(n) = 1 - (1-P)^n 
\end{equation}

Probabilities $P_r$ for the secondary solutions of these 48 outliers are listed in the last column of Table \ref{secondary-solutions}. 34 of them have $P_r$ $<$
5\%, thus confirming the $\zspec$ value with confidence $\gtrsim$ 95\%, while the other 14 have 0.05 $<$ $P_r$ $<$ 0.25, and some of them could be just random alignments.

\subsection{Nature of outliers}

Individualised inspection of outliers reveals a large number of radiogalaxies and radio-loud quasars among them. The remainder are high redshift sources (mostly quasars) with poor SNR spectra.

Among the 21 outliers with no significant ($\gamma$ $>$ 0.01) secondary solutions within $\delta$ $<$ 0.02 of $\zspec$ there are 4 local radiogalaxies (3C 83.1, 3C 465, 3C 371, and 3C 390.3) and two intermediate redshift radio-loud quasars (PG 2251+113 and 3C 295).
The spectra of these 6 sources have high SNR, but are very different from each other. 
3C 390.3 and PG 2251+113 show continuum emission peaking at $\sim$ 20 \uu, a wide silicate emission feature and strong emission in the lines [NeII] 12.81 \uu, [NeIII] 15.55 \uu, and [OIV] 25.91 \uu. 3C 83.1 and 3C 465 have a very weak MIR continuum dominated at $\lambda$ $<$ 10 \um by the Rayleigh-Jeans tail of stellar emission, and also show clear neon lines.
3C 371 is a flat spectrum radio source dominated by synchrotron emission in the MIR with no significant features. Finally, 3C 295 has a steep continuum lacking significant features and seems to have stitching issues in the LL2 module.

The redshift misidentification in all but the last two sources seems not to arise from a lack of spectral features capable of providing an accurate redshift estimate, but rather, from an inadequate representation of these features in the set of templates used. 

Another object, SWIRE J104354.82+585902.4, has conflicting optical redshift estimates: \citet{Trouille08} give $z$ = 0.35, while \citet{Weedman06} give $z$ = 1.14$\pm$0.2, much closer to the value 1.079 found here.

The remaining 14 sources are high redshift ($\zspec$ $>$ 1) quasars and infrared galaxies 
with very low SNR spectra.

\section{Conclusions\label{conclusions}}

In this work we apply a new SED-fitting algorithm to the problem of measuring redshifts in MIR low resolution spectra. The algorithm is based on the same SED-fitting technique applied to broadband photometric redshifts, but with some important modifications that largely increase its efficiency with MIR spectra. Namely: a wavelength dependent scaling factor for the template, which adds flexibility to the fit, and a novel algorithm for filtering and combining prospective redshift solutions, dubbed ``Maximum Combined Pseudo-likelihood'' (MCPL).

The efficiency of MCPL is compared to regular Maximum likelihood (ML) using a sample of 491 \textit{Spitzer}/IRS spectra for sources with accurate optical or NIR spectroscopic redshifts. The spectral templates used are obtained from \textit{Spitzer}/IRS, \textit{AKARI}/IRC and \textit{ISO}/SWS spectroscopy of low redshift galaxies, as well as composite templates of \textit{Spitzer}/IRS spectra of higher redshift sources.

MCPL offers superior performance compared to ML both in terms of the number of highly accurate ($\Delta$($z$)/(1+$z$) $<$ 0.005) redshift solutions (78\% versus 68\% of the sample) and in the number of outliers ($\Delta$($z$)/(1+$z$) $>$ 0.02; MCPL: 14\%, ML: 22\%).
Excluding outliers, the dispersion in the redshift errors is also lower for MCPL: $\sigma$ = 0.0033 (versus 0.0045 for ML).

The reduced $\chi^2$ statistic that determines the goodness of fit, often used to evaluate the reliability of the redshift solution, is found to correlate strongly with the SNR of the spectrum. High SNR spectra obtain higher $\chi^2$ values, indicating the differences between spectrum and template are more evident in them compared to low SNR spectra. Nevertheless, the accuracy of the ML redshift solution does not correlate with $\chi^2$, and thus cannot be directly used to estimate the confidence level of the redshift solution.
On the other hand, the normalised combined pseudo-likelihood ($\gamma$) offers a good indication on the reliability of the MCPL solution for individual spectra, with the median accuracy and rate of outliers both monotonically decreasing with increasing $\gamma$.

The fraction of accurate redshift solutions is much higher among sources classified as starbursts by their MIR emission compared to those classified as AGN (2\% versus 21\% rate of outliers), thanks to the high contrast of the PAH emission bands, which are easily identified even in very low SNR spectra. The rate of outliers is also larger in AGN with the 10 \um silicate feature in emission compared to those in absorption.
Nevertheless, for any given $\gamma$ range the accuracy of MCPL redshifts is largely independent of the MIR SED type.
 
Finally, we find that most outliers are radiogalaxies, radio-loud quasars, or high redshift sources (mostly quasars) with very poor SNR.
About 2/3 of outliers show secondary MCPL solutions at the optical redshift. This indicates that degeneracy issues favoured spurious solutions in the selection process. This could be mitigated with templates that reproduce with greater fidelity the properties of these sources, in particular, radiogalaxies. 

\section*{Acknowledgements}

This work is based on observations made with the \textit{Spitzer Space
Telescope}, which is operated by the Jet Propulsion Laboratory, Caltech
under NASA contract 1407.
The Cornell Atlas of Spitzer/IRS Sources (CASSIS) is a product of the Infrared Science Center at Cornell University, supported by NASA and JPL.
We wish to thank A. Alonso-Herrero, E. Hatziminaoglou, and the anonymous referee for useful comments that helped to improve this paper. 
A.H.-C. is funded by the Universidad de Cantabria Augusto Gonz\'alez Linares program.

\onecolumn

\begin{deluxetable}{l c c c c c c r@{.}l c r@{.}l c} 
\tabletypesize{\small}
\tablecaption{Results for individual sources\label{results-table}}
\tablewidth{0pt}
\tablehead{\colhead{source ID} & \colhead{RA (J2000.0)} & \colhead{Dec (J2000.0)} & \colhead{$\zspec$} & \colhead{z$_{\rm{MCPL}}$} & \colhead{$z_{\rm{ML}}$} & 
\colhead{$\gamma$} & \multicolumn{2}{c}{$R$} & \colhead{$\log \mathscr{N}$} & \multicolumn{2}{c}{$\log \chi^2_{\rm{ML}}$} & \colhead{MIRclass}\\
\colhead{(1)} & \colhead{(2)} & \colhead{(3)} & \colhead{(4)} & \colhead{(5)} & \colhead{(6)} & \colhead{(7)} & \multicolumn{2}{c}{(8)} & \colhead{(9)} & \multicolumn{2}{c}{(10)} & \colhead{(11)}} 
\startdata

                     UGC 00006 &  00:03:09.60 &  +21:57:37.0 &    0.0220 &    0.0222 &    0.0202 &   0.62 &   9 & 83 &   5.30 &   2 & 79 &     MIR\_SB\\
             {[}HB89] 0003+199 &  00:06:19.50 &  +20:12:10.0 &    0.0260 &    0.0284 &    0.0284 &   0.20 &   2 & 19 &   3.75 &   0 & 20 &   MIR\_AGN1\\
       2MASX J00070361+1554240 &  00:07:03.60 &  +15:54:24.0 &    0.1140 &    0.1118 &    1.5595 &   0.15 &   1 & 44 &   4.48 &   1 & 64 &   MIR\_AGNx\\
                    III Zw 002 &  00:10:30.80 &  +10:58:13.0 &    0.0898 &    0.0898 &    0.0876 &   0.17 &   2 & 45 &   4.79 &   1 & 99 &   MIR\_AGN1\\
                      NGC 0017 &  00:11:06.50 &  -12:06:26.0 &    0.0196 &    0.0202 &    0.0202 &   0.52 &  19 & 96 &   4.11 &   1 & 73 &     MIR\_SB\\
       2MASX J00114330-0722073 &  00:11:43.30 &  -07:22:07.0 &    0.1180 &    0.1185 &    0.1185 &   0.82 &  17 & 71 &   4.99 &   2 & 02 &     MIR\_SB\\
       2MASX J00212652-0839261 &  00:21:26.50 &  -08:39:26.0 &    0.1280 &    0.1320 &    0.1320 &   0.46 &   9 & 23 &   4.97 &   2 & 48 &   MIR\_AGN2\\
                LBQS 0018-0220 &  00:21:27.30 &  -02:03:33.0 &    2.5960 &    0.8183 &    2.5958 &   0.11 &   1 & 80 &   3.60 &   0 & 31 &    MIR\_AGN\\
\enddata
\tablecomments{(1) Source identification in NED. (2),(3) J2000 right ascension and declination. 
(4) Spectroscopic redshift from NED. (5),(6) MCPL and ML redshift solutions, respectively. (7) Reliability
parameter for MCPL solution. (8) Ratio of $\gamma$ values for the first and second MCPL redshift solutions. (9) Logarithm of the net significance.
(10) Logarithm of the reduced $\chi^2$ for the ML solution.  (11) MIR classification of the source.}

\end{deluxetable}

\begin{deluxetable}{r l r@{.}l l l} 
\tabletypesize{\small}
\tablecaption{Infrared templates\label{table-templates}}
\tablewidth{0pt}
\tablehead{\colhead{No.} & \colhead{template ID} & \multicolumn{2}{c}{redshift} & \colhead{type} & \colhead{source}}
\startdata
1 & NGC 5011         &  0 & 00105& El-2           & CWW\_E + IRS\tablenotemark{a} \\
2 & PG 1116+215      &  0 & 1765 & QSO (S1.0)     & QSO\_comp + IRS \\
3 & PG 1048+342      &  0 & 16713 & Elliptical (S1.0) & QSO\_comp + IRS\tablenotemark{b} \\
4 & 2MASS J12324114+1112587 & 0 & 249 & NLAGN & QSO\_comp + IRS \\
5 & 2MASXi J2222211+195947 & 0 & 21 & Seyfert (S1.0) & IRS + IRS\tablenotemark{c} \\
6 & PG 1440+356     &  0 & 079  & QSO (S1n)      & QSO\_comp + IRS \\
7 & [HB89] 1415+451 &  0 & 1136 & RQQ (S1.0)     & QSO\_comp + IRS \\
8 & 3C 120          &  0 & 03301 & BLRG   & QSO\_comp + IRS \\
\enddata
\tablenotetext{a}{Optical-NIR elliptical galaxy template from \citet{Coleman80}.}
\tablenotetext{b}{ Glikman + blueAGN composite.}
\tablenotetext{c}{Extended at shorter wavelengths using IRS spectrum of BAL QSO ELAISC15 J003059-442133 ($z$=2.101).}
\tablenotetext{d}{composite spectrum of radiogalaxies with strong emission in the [SIV] 10.51 \um line.}

\end{deluxetable}

\begin{deluxetable}{c c c c c c c} 
\tabletypesize{\small}
\tablecaption{Reliability of redshift meassurements\hfill\label{reliability-table}}
\tablewidth{0pt}
\tablehead{\colhead{$\gamma$ range} & \colhead{N sources} & \colhead{$\sigma$($d$)\tablenotemark{a}} & \colhead{median $\delta$\tablenotemark{b}} &
\colhead{$f$($\delta$ $<$ 0.02)\tablenotemark{c}} & \colhead{$f$($\delta$ $>$ 0.02)\tablenotemark{d}} & \colhead{$f$($\delta$ $>$ 0.1)\tablenotemark{e}}}
\startdata
0.05--0.1 &  73 &   0.0279 &   0.0040 &   0.53 &   0.47 &   0.30\\
0.1--0.15 &  80 &   0.0204 &   0.0020 &   0.73 &   0.28 &   0.12\\
0.15--0.2 &  49 &   0.0131 &   0.0019 &   0.88 &   0.12 &   0.08\\
 0.2--0.3 &  89 &   0.0084 &   0.0011 &   0.96 &   0.04 &   0.03\\
 0.3--0.5 & 123 &   0.0056 &   0.0009 &   0.98 &   0.02 &   0.01\\
 0.5--1.0 &  77 &   0.0020 &   0.0009 &   1.00 &   0.00 &   0.00\\
\enddata
\tablenotetext{a}{standard deviation of redshift errors (excludes catastrophic errors).}
\tablenotetext{b}{median redshift accuracy (excludes catastrophic errors).}
\tablenotetext{c}{frequency of accurate redshift solutions.}
\tablenotetext{d}{frequency of outliers.}
\tablenotetext{e}{frequency of catastrophic errors.}

\end{deluxetable}

\begin{deluxetable}{l r r r r r} 
\tabletypesize{\small}
\tablecaption{Dependency with MIR type\label{MIRtype-table}}
\tablewidth{0pt}
\tablehead{\colhead{subset} & \colhead{MIR\_SB} & \colhead{MIR\_AGN} & \colhead{MIR\_AGN1} & \colhead{MIR\_AGN2} & \colhead{Total}}
\startdata
                            $\delta$ $<$ 0.02 & 179 & 223 &  80 & 114 & 422\\
                          $\delta$ $\ge$ 0.02 &   2 &  62 &  33 &  10 &  69\\
                            $\gamma$ $>$ 0.15 & 175 & 149 &  54 &  86 & 338\\
       $\gamma$ $>$ 0.15 \& $\delta$ $<$ 0.02 & 173 & 139 &  47 &  85 & 325\\
       $\gamma$ $>$ 0.15 \& $\delta$ $\ge$ 0.02 &   2 &  10 &   7 &   1 &  13\\
                                        Total & 181 & 285 & 113 & 124 & 491\\
\enddata
\end{deluxetable}

\begin{deluxetable}{l r r r r@{.}l r r r}
\tabletypesize{\small}
\tablecaption{Sources with accurate secondary redshift solutions\label{secondary-solutions}}
\tablewidth{0pt}
\tablehead{\colhead{source name} & \colhead{$\zspecu$} & \colhead{$z_{\rm{MCPL}}$} & \colhead{$z_{\rm{alt}}$} &
\multicolumn{2}{c}{$d_{\rm{alt}}$} & \colhead{$\gamma_{\rm{alt}}$} & \colhead{\#} & \colhead{P$_r$} \\
\colhead{(1)} &\colhead{(2)} &\colhead{(3)} &\colhead{(4)} &\multicolumn{2}{c}{(5)} &\colhead{(6)} &\colhead{(7)} &\colhead{(8)}}
\startdata
{[}HB89{]} 1435-067           &    0.1260 &    0.0768 &    0.1343 &    0 & 0073 &    0.017 & 20 &    0.2059\\
{[}HB89{]} 1543+489           &    0.3996 &    0.8662 &    0.4077 &    0 & 0058 &    0.052 &  6 &    0.0584\\
2MASSi J1516532+190048        &    0.1900 &    0.2386 &    0.1924 &    0 & 0020 &    0.053 &  6 &    0.0265\\
2MASSi J1649149+530316        &    2.2600 &    2.5317 &    2.2084 &   -0 & 0158 &    0.081 &  2 &    0.0390\\
2MASX J00215355-7910077       &    0.0728 &    0.0304 &    0.0725 &   -0 & 0003 &    0.090 &  2 &    0.0020\\
2MASX J01500266-0725482       &    0.0180 &    0.0876 &    0.0182 &    0 & 0002 &    0.186 &  3 &    0.0012\\
2MASX J14025121+2631175       &    0.1870 &    0.1185 &    0.1641 &   -0 & 0193 &    0.071 &  5 &    0.1313\\
3C 015                        &    0.0730 &    3.3134 &    0.0725 &   -0 & 0005 &    0.045 & 13 &    0.0076\\
\enddata
\tablecomments{(1) Source identifier in NED. (2) Optical spectroscopic redshift. (3) MCPL redshift solution. (4) Redshift of secondary 
MCPL solution that matches $\zspec$. (5) Redshift error for secondary solution. (6) Reliability parameter for secondary solution. (7) Order
number of the secondary solution after sorting by decreasing $\gamma$ values. (8) Probability of secondary solution being spurious.}
\end{deluxetable}

\end{document}